\documentstyle[epsfig,aps,pra]{revtex}
\begin{document}

\bibliographystyle{unsrt}

\title{1S-2S Spectrum
of a Hydrogen Bose-Einstein Condensate\\
{\it published Physical Review A {\bf 61}, 33611 (2000)}
}

\author{Thomas C. Killian\cite{tk}}

\address{
Department of Physics and Center for Materials Science and
Engineering, Massachusetts Institute of Technology, Cambridge,
Massachusetts 02139
}

\maketitle

\begin{abstract}
We calculate the  
two-photon $1S$-$2S$ spectrum of an atomic hydrogen 
Bose-Einstein condensate in the regime where
the cold collision frequency shift 
dominates the lineshape.
WKB and static phase
approximations are made to find the intensities 
for transitions from the 
condensate  to  motional
eigenstates for  $2S$ atoms.
The excited state wave functions are found using a mean field potential
which includes the effects of collisions with condensate atoms.
Results agree well  with experimental data. This formalism can be used
to find condensate spectra for a wide range of  
excitation schemes.

\end{abstract}

\section{Introduction}
In the recent experimental observation of Bose-Einstein condensation 
(BEC)
in atomic hydrogen \cite{fkw98}, the cold collision frequency
shift in the $1S$-$2S$ photoexcitation spectrum \cite{kfw98} signalled
the presence of a condensate.
The shift arises because electronic
energy levels are perturbed
due to interactions, or collisions, with neighboring atoms.
In the cold collision regime, 
 the temperature  is low enough that
the $s$-wave scattering length, $a$,
is much less than the  thermal de 
Broglie wavelength,
$\lambda_{T}=\sqrt{h^2/2\pi mk_B T}$, and
only $s$-waves are involved in the collisions\cite{jmi89}.

The cold
collision frequency shift has also been studied
in the hyperfine spectrum of hydrogen in  cryogenic masers \cite{kcs88},
and 
cesium \cite{tvs92,gch93,gll96}
and rubidium \cite{kvg97} in  
atomic fountains.
Theoretical explanations of these results and
other work on the  hydrogen $1S$-$2S$
spectrum \cite{ole99,okk99}
have focused on the magnitude of the 
shift, as opposed to a line{\it shape}.
In this article 
 we  present a calculation of the hydrogen BEC $1S$-$2S$ spectrum.
We also describe how the formalism can be used for other
atomic systems and experimental conditions.


\subsection{The Experiment}
The experiment is described in \cite{fkw98,kfw98}, and we summarize the 
important aspects here.
Hydrogen atoms in the $1S$, $F=1$, $m_{F}=1$ state are
confined in a  magnetic
trap and evaporatively cooled.
The hydrogen condensate
is observed in the temperature range 
$30$-$70$~$\mu$K and the condensate fraction never
exceeds a few percent. Nevertheless, 
the peak density in the normal cloud
is almost two orders of magnitude lower than in the condensate and
in this study we will neglect the presence of the noncondensed gas.

The two-photon transition to the 
metastable $2S$, $F=1$, $m_{F}=1$ state ($\tau=122$~ms) is driven by
a 243~nm laser beam which passes through the sample and
is retroreflected. In this
configuration, an atom can absorb one photon from each
direction. This results in Doppler-free excitation  for
which there is no momentum transferred to the
atom and no Doppler-broadening of the resonance. An
atom can also absorb two co-propagating photons and
receive a momentum kick. This  is    
Doppler-sensitive  excitation, and the spectrum  in this case is
recoil shifted and Doppler-broadened.
The photo-excitation rate is monitored by 
counting $122$~nm fluorescence photons from the excited state.
For a typical laser pulse of $500$~$\mu$s,
fewer than 1 in $10^{4}$ of the atoms are promoted to the $2S$ state.
$2S$ atoms experience the same trapping potential as $1S$ atoms because the 
magnetic
moment is the same for both states, neglecting small relativistic 
corrections.

The natural linewidth of the $1S$-$2S$  
transition is $1.3$~Hz, but the 
experimental width, 
at low density and temperature, is limited by the laser coherence time.
The narrowest observed spectra, obtained when studying a noncondensed
gas, have widths of  a few
kHz \cite{cfk96}.
For the   condensate, 
the cold collision frequency
shift is as much as one MHz and it dominates the lineshape.

\subsection{Mean Field Description of the Spectrum}
The  frequency shift in maser and  fountain 
experiments has traditionally been 
described using the 
quantum Boltzmann equation\cite{kcs88,tvs92,kvg97}. In this picture,
the frequency shift is the net result of the  small
collisional phase shifts arising from forward scattering events in the gas.
A mean field description, however,  is more convenient
for studying an inhomogeneous  Bose-Einstein condensate. 
We will derive this 
picture in detail, but we summarize the results here.
Collisions add a mean field energy to  the atom's  potential energy. 
For a $2S$ atom excited 
out of a condensate  the mean field term is
$\delta E_{2S}({\bf r})=4\pi \hbar^2 a_{1S-2S}n_{1S}({\bf r})/m$. 
For a  $1S$ condensate atom the mean field term is
$\delta E_{1S}({\bf r})=4\pi \hbar^2 a_{1S-1S}n_{1S}({\bf r})/m$. 
(The fraction of excited $2S$ atoms is small, so
$2S$-$2S$ interactions can be neglected.)
The ground state $s$-wave triplet scattering length  has been
calculated accurately 
($a_{1S-1S}=0.0648$~nm\cite{jdk95}). The $1S$-$2S$ scattering length,
however,
is less well known 
($a_{1S-2S}=-1.4\pm0.3$~nm from experiment \cite{kfw98} and 
 -2.3~nm from theory \cite{jdd96}).

\begin{center}

\begin{figure}
\epsfig{file=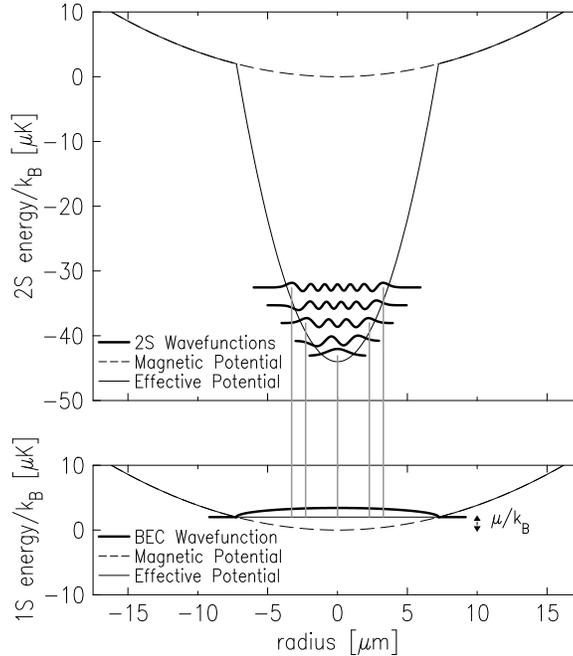, width=3in}
\caption
{Effective potentials for $1S$ atoms in the
condensate and excited $2S$ atoms. 
Selected single particle 
wave functions are  displayed at the height corresponding to their
energy. The
dashed lines are the magnetic trapping potential $V({\bf r})$, which
is identical for $1S$ and $2S$ atoms. The thin solid lines
are the effective potentials, which include
the mean field interaction energy. 
The vertical light solid lines indicate
allowed Doppler-free transitions from the condensate, which
must preserve mirror symmetry. The  
potentials and condensate wave function
are for a peak condensate density of 
$5 \times 10^{15}$ cm$^{-3}$
and a magnetic
trap oscillation frequency of 4 kHz, 
which are characteristic conditions for a
hydrogen BEC and a strong confinement axis of the trap [1,2]. 
The scattering lengths used in the calculations
are $a_{1S-1S}=0.0648$~nm and $a_{1S-2S}=-1.4$~nm,
and the chemical potential is
$\mu/k_B \approx 2~\mu$K. 
The $2S$ levels form a near continuum of 
motional states in an 
anisotropic three dimensional trap. }
\label{veffbec}
\end{figure}

\end{center}

We denote
the sum of the magnetic trap potential, $V({\bf r})$, 
and the mean field  energy, $\delta E_{x}({\bf r})$,
as the effective potential, $V_{x}^{eff}({\bf r})$ (Fig.\ \ref{veffbec}).
Here $x$ is either $1S$ or $2S$. For
$1S$ condensate atoms, the  effective potential in the condensate is flat.
Because $a_{1S-2S}<0$ and the condensate density is large, $2S$ atoms 
experience a stiff attractive potential in the condensate
which  supports many bound $2S$ motional states.

The $1S$-$2S$  spectrum consists of
transitions from the condensate to $2S$
motional eigenstates of the
effective $2S$ potential.
For Doppler-free excitation, the final states are bound in the
BEC well. Doppler-sensitive excitation populates states
which lie about $\hbar^{2}k_{0}^{2}/2m k_{B} =  643$~$\mu$K above
the bottom of the $2S$
potential, where $\hbar k_{0}$ is the momentum carried by two 
laser photons. The latter states extend over a region much greater than the 
condensate. Because the excited levels are so different for 
Doppler-free and Doppler-sensitive excitation, we must treat the two
spectra independently.

The rest of this article
presents  a derivation of the effective potentials
and a quantum mechanical  calculation of the BEC
$1S$-$2S$  spectrum.

\section{1S-2S Photoexcitation Spectrum of a 
Hydrogen Bose-Einstein Condensate}
\label{Ainhomosec}

\label{cstbecspectrum}

\subsection{Hamiltonian}
We start with the many-body Hamiltonian for a system with $N$ 
atoms,
\begin{eqnarray}
H=\sum_{j=1}^N\left({{p}_j^2 \over 2m} + 
H^{int}_j + V({\bf { r}}_j)\right) + H^{las} + H^{coll}, \label{Htrap}
\end{eqnarray}
where ${\bf { p}}_j$, ${\bf { r}}_j$, and 
$H^{int}_j$ are  the momentum operator, 
position operator, and internal state Hamiltonian
respectively
for particle $j$.
$V({\bf r})$ is the magnetic trapping potential,
which is 
the same for $1S$ and $2S$ atoms.

$H^{las}$ is the atom-laser interaction.
After making the rotating wave approximation, it can be written 
\begin{equation}
H^{las}=\sum_{j=1}^N {\hbar \Omega({\bf {r}}_j)
\over
2}
\left[(|2S\rangle\langle 1S|)_j e^{-i4\pi \nu t} +(|1S\rangle\langle 
2S|)_j e^{i4\pi \nu t}\right], \label{laserham}
\end{equation}
where $\nu$ is the frequency of the laser field ($2h\nu \approx 
E_{2S}-E_{1S}\equiv E_{1S-2S}$
on resonance).
The laser beam is uniform over the condensate, so
we treat the excitation as 
a standing wave consisting of two counter-propagating plane waves.
The effective 2-photon Rabi frequency 
for  Doppler-free  excitation \cite{bha86},
\begin{equation}
\Omega_{DF}({\bf { r}}) = \Omega_{DF}= {2M_{2S,1S} \over 3 \pi^2 \hbar c}
\left({\alpha \over 2 
R_{\infty}}\right)^3 I ,
\end{equation}
 is uniform in space. Here, $I$ is
the laser intensity in each direction,
$M_{2S,1S}=11.78$  \cite{bfq77} is a unitless constant,
$c$ is the speed of light in vacuum, $\alpha$ is the fine structure
constant, and $R_{\infty}$ is the Rydberg constant.
For Doppler-sensitive excitation,
\begin{equation}
\Omega_{DS}({\bf { r}}) =  \Omega_{DS}({\rm e}^{ik_{0}¥z}+ {\rm 
e}^{-ik_{0}¥z}),\label{dsrabi} 
\end{equation}
where $\Omega_{DS}=\Omega_{DF}/2$.

$H^{coll}$ 
describes the effects of two-body elastic
collisions. 
In the cold collision regime, the interaction can be represented by a
shape independent pseudopotential\cite{hua87} corresponding to a
phase shift per collision of $ka$, where $\hbar {k}$ is the momentum
of each of the colliding particles in the center of mass frame,
\begin{eqnarray}
H^{coll}&=& {4\pi \hbar^2 \over m}\sum_{i<j}^N\delta ({\bf { r}}_i - 
{\bf { r}}_j)
\left[a_{1S-1S} (|1S\rangle \langle 1S|)_i(|1S\rangle \langle 1S|)_j 
+ a_{1S-2S}(|e^{3}\Sigma^{+}_{u}\rangle \langle 
e^{3}\Sigma^{+}_{u}|)_{ij}\right.\nonumber \\
&&\left.+ a_{2S-2S} (|2S\rangle \langle 2S|)_i(|2S\rangle \langle 2S|)_j\right]. 
\label{AinteractionH}
\end{eqnarray}
The sum is over $N(N-1)/2$ distinct pairwise interaction
terms. The $1S$-$2S$ interaction projection operator is written in 
terms of 
\begin{eqnarray}
|e^{3}\Sigma^{+}_{u}\rangle_{ij}&=& 
{|1S\rangle _i|2S\rangle_j +|2S\rangle _i|1S\rangle_j\over \sqrt{2}}
\end{eqnarray}
because the doubly spin polarized atoms collide on the 
$e^{3}\Sigma^{+}_{u}$ potential during $s$-wave collisions \cite{jdd96}.
As mentioned above, the $2S$-$2S$ scattering term is negligible
for the hydrogen experiment,
but it is included here for completeness.

Inelastic collisions, such as collisions in which the hyperfine level of
one or both of the colliding partners changes, will contribute additional
shifts which are not included in this formalism, but 
these effects are expected to be small in the experiment \cite{kfw98}. 

\subsection{System before Laser Excitation}
\label{condsec}
We make the approximation that  the system is at
$T=0$, and
all  atoms are initially in the condensate.
$T=0$ models have accurately 
described many condensates properties\cite{dgp99}, and 
we leave finite temperature
effects for  future study. 
The state vector can be written 
\begin{equation}
|\Psi_{0}\rangle = |\underbrace{1S, 0;...;1S,0}_{\rm 
N~terms}\rangle. \label{define}
\end{equation}
where $|1S,0\rangle$ refers to the single particle electronic and
motional state of an atom in a $1S$ condensate with N atoms.
We use the ket  notation ($|a;b;..;c\rangle$), in which
the entry in the first slot is the state of atom 1, the second entry
is the state of atom 2, {\it etc}.

Minimization of
$\langle \Psi_{0}|H |\Psi_{0}\rangle$ 
leads to the Gross-Pitaevskii, or nonlinear
Schr\"{o}dinger equation \cite{gpi58,gro63} for the single particle
BEC wave function,
$\psi({\bf r})= \langle {\bf r}|0\rangle$, 
\begin{equation}
\mu \psi ({\bf r}) =\left( - {\hbar^{2}¥\nabla^{2}¥\over 2m} + V_{1S}^{eff}({\bf r})
\right)\psi({\bf r}).
\end{equation}
The effective potential is
$V_{1S}^{eff}({\bf r})=V({\bf r}) + \tilde{U} n({\bf r})$,
where $\tilde{U} =4\pi \hbar^2 a_{1S-1S} /m$.
Here, $n({\bf r})=N|\psi({\bf r})|^{2}$ is the density
distribution in the N-particle condensate.
One can interpret
$|\psi({\bf r}_i)|^{2}$ as the probability
of finding condensate particle $i$ at position ${\bf r}_i$. 

The kinetic energy is small and can be neglected.
This yields the Thomas-Fermi 
wave function \cite{bpe96}, 
\begin{eqnarray}
\psi({\bf r}) 
&=&\left\{ \begin{array}{ll}
N^{-1/2} \left[ n(0) - V({\bf r})/\tilde{U} \right]^{1/2} & 
\mbox{$V({\bf r}) \leq n(0)\tilde{U}$} \\
0								& \mbox{otherwise}
\end{array} \right. ,\label{define1}
\end{eqnarray}
where 
$n(0)$
is the peak density.
The density
profile is the inverted image of 
the trapping potential.
The chemical potential is
$\mu(N)= 
\tilde{U}n(0)$, and it is equal to $V_{1S}^{eff}$ inside the
condensate.
The energy of the system before laser excitation
is the minimum of $\langle \Psi_{0}|H |\Psi_{0}\rangle$. It
satisfies  $\mu(N)=\partial E_{0} /\partial N$ and is given by
\begin{equation}
E_{0}={5\over7}N \mu (N). \label{NatomBECenergy}
\end{equation}
From now on, when writing  $\mu$
we will drop the explicit dependence on N.
For 
a cylindrically symmetric harmonic trap, it can be shown that 
$
n(0)= \left(15Nm^3w_r^2w_z/\hbar^3a^{3/2}_{1S-1S}\right)^{2/5}/8\pi, 
$
where $w_r$ and $w_z$ are the angular frequencies for
 radial and axial oscillations in the trap. 

\subsection{System after Laser Excitation
}

To describe the system after laser excitation we must find  
the  orthonormal basis of $2S$ motional wave functions and
their energies. This is done by minimizing 
$\langle \Phi_{q,i}|H 
|\Phi_{q,i}\rangle$, where 
\begin{eqnarray}
|\Phi_{q,i}\rangle &=& \hat{\cal 
S}|\underbrace{2S,i;...;2S,i}_{\rm 
q~terms};\underbrace{1S,0;...;1S,0}_{\rm N-q~terms} 
\rangle \label{excitedbra}
\end{eqnarray}
is a  state with $q$ $2S$ atoms in $2S$ motional level $i$.
The operator $\hat{\cal S}$ symmetrizes  with
respect to particle label.
We will show below that the state vector of the system after laser
excitation is actually expressed as a 
superposition of such terms, but for now we
need only consider a single $|\Phi_{q,i}\rangle$.

Calculating $\langle \Phi_{q,i}|H 
|\Phi_{q,i}\rangle$ involves a somewhat lengthy calculation.
Details are given in appendix \ref{excbectrap} and  the result 
is
\begin{eqnarray}
\langle \Phi_{q,i}|H 
|\Phi_{q,i}\rangle&=& E'_{0} \nonumber \\
& & + q\, \langle 2S, i|
\left[ H^{int} +{{ p}^2 \over 2m}  + V_{2S}^{eff}({\bf r}) \right]
|2S,i\rangle \nonumber \\
&=&E'_{0} + q(E_{1S-2S} + \varepsilon_i).\label{some2SBECenergy}
\end{eqnarray}
$E'_{0}$ is the energy of a pure $1S$ condensate with $N-q$ atoms
(see Eq.\ \ref{NatomBECenergy} and \ref{eprime}),
$\varepsilon_i=\langle i|
\left[{{ p}^2 \over 2m}  + V_{2S}^{eff}({\bf r}) \right]
|i\rangle$,
 and
the effective potential for the $2S$ atoms is 
\begin{equation}
V_{2S}^{eff}({\bf r}) = V({\bf {\bf r}})+  
{4\pi\hbar^2a_{1S-2S} \over m} \, n_{N-q}({\bf r}). \label{2seff}
\end{equation} 
The density of $1S$ atoms remaining is $n_{N-q}({\bf r})=(N-q)|\psi({\bf 
r})|^{2}$.  
 
Finding the $2S$ motional states which minimize $\langle \Phi_{q,i}|H 
|\Phi_{q,i}\rangle$, with the 
requirement
that  they form an orthonormal basis,
is equivalent to finding the eigenstates of the effective $2S$ Hamiltonian
\begin{eqnarray}
H_{2S}^{eff} = {{ p}^2 \over 2m} + V_{2S}^{eff}({\bf r}), 
\label{localHam}
\end{eqnarray}
and the  eigenvalue for state $i$ is $\varepsilon_i$. The effective Hamiltonian
(Eq.\ \ref{localHam}) is consistent with the two-component Hartree-Fock 
equations used to calculate the single particle wavefunctions for
double condensates \cite{egb97}.
The effective potential and some $2S$ motional states are  depicted in 
Fig.\ \ref{veffbec}. 

If we denote the minimum of $\langle \Phi_{q,i}|H 
|\Phi_{q,i}\rangle$ as $E_{q,i}$,
using Eq.\ \ref{NatomBECenergy} and \ref{some2SBECenergy}, 
the energy supplied by two photons to drive the transition to state
$i$, for $q \ll N$,  is 
\begin{eqnarray}
2h\nu&=&{E_{q,i}-E_{0} \over q}={q(E_{1S-2S} + 
\varepsilon_i) +E'_{0}-E_{0} \over q}\nonumber \\
&\approx& E_{1S-2S} + \varepsilon_i -\mu.
\end{eqnarray}
We have used $(E_{0}-E'_{0}) /q \approx \partial 
E_{0} /\partial N= \mu$ for small $q$. Note that $\varepsilon_i 
<0$ for
states bound in the BEC interaction well.
Since many $2S$ motional levels  may be excited, there will 
be a distribution
of excitation energies in the spectrum.

When condensate atoms are coherently excited to  an isolated level $|i\rangle$ by
a laser pulse of duration $t$,
the single particle wave functions evolve according to  \cite{mak97}
\begin{equation}
|1S,0\rangle  \Rightarrow {\rm cos}\theta|1S,0\rangle + 
{\rm sin}\theta|2S,i\rangle, \label{rotate}
\end{equation}
where
\begin{equation}
{\rm sin}^{2}\theta={|\langle i|\Omega({\bf r})|0 \rangle|^{2} \over 
|\langle i|\Omega({\bf r})|0 \rangle|^{2}+\delta \omega^{2}}
{\rm sin}^{2}\left[\left(|\langle i|\Omega({\bf r})|0 \rangle|^{2} 
+\delta \omega^{2}\right)^{1/2}t/2\right]. \label{sin2}
\end{equation}
The detuning from resonance is $\delta\omega$.
In Eq.\ \ref{rotate}, we assume  the excitation is  weak enough to 
neglect the change in the
single particle wave function for atoms in the condensate\cite{mhj98,hme98}.
Depending upon which excitation scheme  is being described,
$\Omega({\bf r})$ is either $\Omega_{DF}¥({\bf r})$ 
or $\Omega_{DS}¥({\bf r})$.

The state vector for the system after excitation can be written
\begin{eqnarray}
|\Psi_{{\langle q\rangle},i}\rangle &=& 
\underbrace{({\rm cos}\theta|1S,0\rangle + 
{\rm sin}\theta|2S,i\rangle) 
\otimes \ldots \otimes
({\rm cos}\theta|1S,0\rangle + 
{\rm sin}\theta|2S,i\rangle)
}_{\rm N~terms}\nonumber \\
&=&\sum_{q=0}^N 
{\rm cos}^{N-q}\theta\, {\rm sin}^q\theta \sqrt{N! \over 
q!(N-q)!}|\Phi_{q,i}\rangle,  \label{bigexcited}
\end{eqnarray}
where the label $\langle q\rangle = N{\rm sin}^{2}\theta$ is the expectation 
value of the  number
of $2S$ atoms excited. Although $q$ is not a good quantum
number for $|\Psi_{{\langle q\rangle},i}\rangle$, the spread 
in $q$, given by a binomial distribution, is strongly peaked
around $\langle q\rangle$.

For short excitation times, the population in state $i$  
grows
coherently as $t^{2}$.  
For the hydrogen experiment, however, although the excitation is weak 
and
$|\langle i|\Omega({\bf r})|0\rangle|\, t\ll 1$,
$t$ is  longer than the
coherence time of the laser ($\sim 200$~$\mu$s). 
This implies that the  number of atoms 
excited to level $i$ must be  expressed in a form
reminiscent of 
Fermi's Golden Rule. 
Equation \ref{sin2} can be rewritten in terms of a delta function using
the relation
$\rm{sin}^{2}(xt)/\pi x^{2}t \rightarrow 
\delta(x)$ as $t\rightarrow \infty$.
(One can neglect $|\langle i|\Omega({\bf r})|0\rangle|$ compared
to $\delta\omega$ because $|\langle i|\Omega({\bf r})|0\rangle|$ is 
small compared to the spread in frequency of the laser excitation.) Then
\begin{eqnarray}
\langle q\rangle 
&\approx&{N \pi \hbar t\over 2}
 |\langle i|\Omega({\bf r})|0\rangle|^2\,  
\delta(2h\nu-E_{1S-2S} - \varepsilon_i + \mu). \label{onelevel}
\end{eqnarray}
It is understood that Eq.\ \ref{onelevel} is to be convolved with
the laser spectrum or a density of states function.
The total $2S$ excitation rate is
\begin{eqnarray}
S(2h\nu) &=&{N \pi \hbar \over 2}
\sum_{i} |\langle i|\Omega({\bf r})|0\rangle|^2\,  
\delta(2h\nu-E_{1S-2S} - \varepsilon_i + \mu) \nonumber \\
&=&{N \pi \hbar \Omega^{2}\over 2}
\sum_{i} F^{i}\,  
\delta(2h\nu-E_{1S-2S} - \varepsilon_i + \mu). 
\label{overlapint} 
\end{eqnarray}
Equation \ref{overlapint} defines the overlap factors,
$F^{i}=|\langle i|\Omega({\bf 
r})/\Omega|0\rangle|^2$, which
are analogous to Franck-Condon factors in molecular spectroscopy.
An expression equivalent to Eq.\ \ref{overlapint}, the
strength distribution function or dynamic form factor, is
commonly used to describe  collective excitations of many body 
systems \cite{dgp99}. 

The BEC spectrum now appears as $N$ times the spectrum of
a  single particle in $|0\rangle$ excited to
eigenstates of the effective $2S$ potential.
The broadening in
the $1S$-$2S$ BEC spectrum is homogeneous because it results from
a spread in the energy of possible excited states, not from a spread 
in the energy of initially occupied states.

The central results of this calculation are the effective
$2S$ potential (Eq.\ \ref{2seff}) and the Fermi's Golden Rule 
expression for the excitation rate (Eq.\ \ref{overlapint}).
Using this formalism we can now calculate the observed spectrum for 
Doppler-free and Doppler-sensitive excitation.

\subsection{Doppler-Free 1S-2S Spectrum}
Doppler-free excitation populates states which
are bound inside the BEC potential well (see Fig.\ \ref{veffbec}). 
For a condensate in a 
harmonic trap, these states
are approximately eigenstates of a three dimensional harmonic oscillator
with trap frequencies larger than those of
the magnetic trap alone by a factor of 
$\sqrt{1-a_{1S-2S}/a_{1S-1S}}\approx 5$ (see Eq. \ref{define1} and 
\ref{2seff}). 
Because  we know the wave functions, we can numerically evaluate
Eq.\ \ref{overlapint}. 
The result of such a calculation 
is shown in Fig.\ \ref{overlap}.

\begin{center}
\epsfig{file=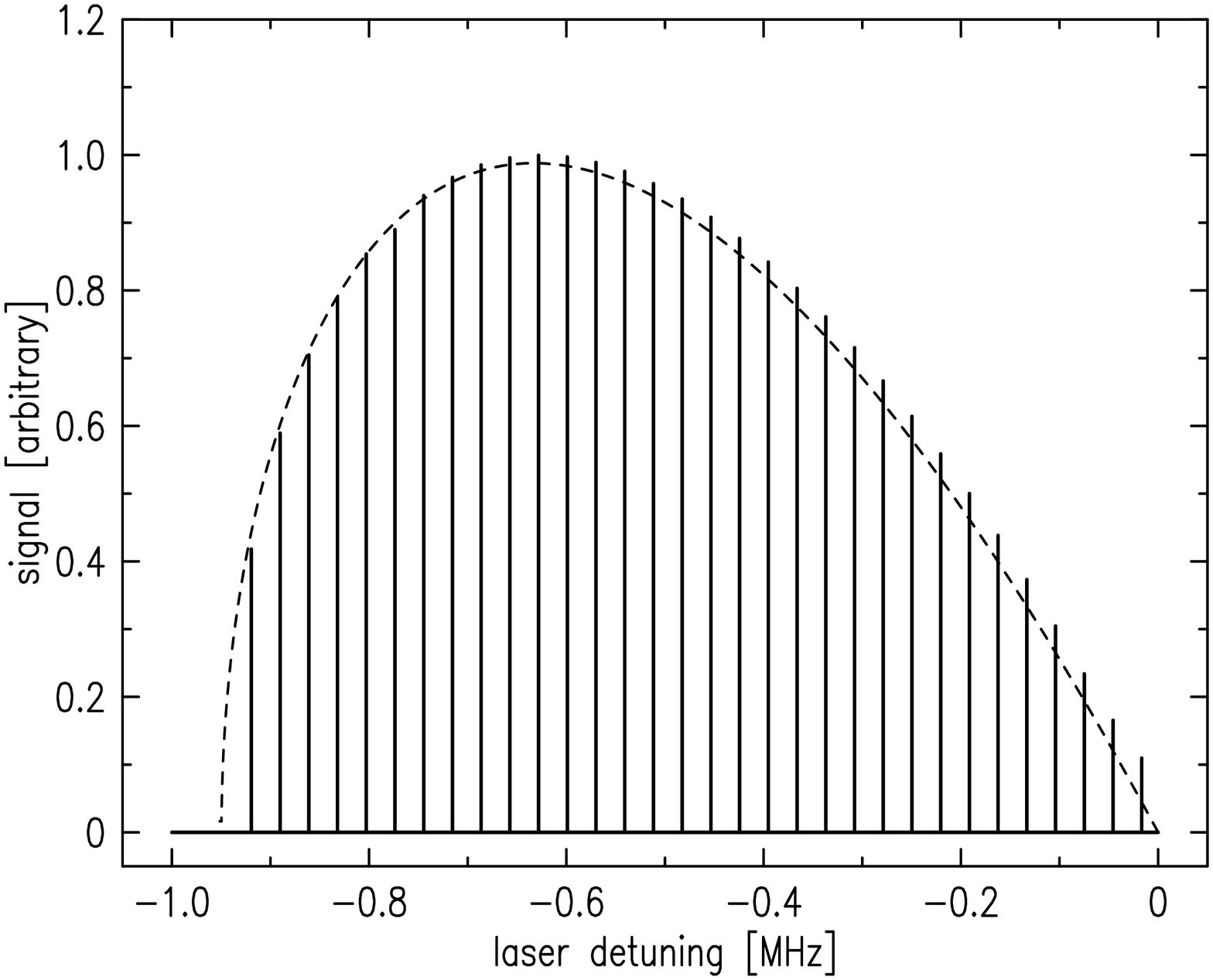, width=3in}

\begin{figure}
\caption
{Calculated Doppler-free spectrum of a condensate at $T=0$ in a 
three-dimensional harmonic trap. Zero detuning is the unperturbed
Doppler-free transition frequency. 
The stick spectrum results from the sum
over the transition amplitudes expressed in Eq.\ \ref{overlapint}
using the Thomas-Fermi density 
distribution for a peak
condensate density of $10^{16}$~cm$^{-3}$
($4\pi\hbar^2 (a_{1S-2S}-a_{1S-1S}) n({0})/ m\approx 2h \times -0.95$~MHz).
The trap is spherically symmetric  with $\omega_{trap}=2\pi \times 6$~kHz.
The stick heights represent the coefficients
of delta functions which must be convolved with the laser spectrum
of about $1\, $kHz FWHM.
The dashed curve (Eq.\ \ref{reduce}) follows from 
the integral over the BEC density distribution, 
Eq.\ \ref{radialintegral}, 
for the same peak condensate density.  The envelope is independent
of the symmetry of the trap, but the stick spectrum blends into
a continuum in a trap with one weak confinement axis such as the MIT
hydrogen trap [1,2]. Resolution of the individual transitions would 
require a stiff, near spherically symmetric trap, very stable
experimental conditions, and high signal/noise. It does not seem 
feasible with the hydrogen experiment in the near future.}
\label{overlap}
\end{figure}

\end{center}

At large red detuning ($2h\,\delta\nu \approx  4\pi\hbar^2 a_{1S-2S} n(0)/m - 
\mu$) transitions are to the lowest state in the 
BEC interaction
well.
The spectrum does not extend  to the blue of $2h\,\delta\nu = 0$
because states outside the well have 
negligible overlap with the condensate
and are  inaccessible by laser excitation.
In the overlap integrals in Fig.\ \ref{overlap}, wave functions for an
infinite harmonic trap were used for the $2S$ motional states. These 
deviate from the
actual motional states near the top of the BEC interaction well, 
introducing small errors in the stick spectrum
nearer zero detuning. 

The envelope of the spectrum in Fig.\ \ref{overlap}
can be derived analytically and
reveals some interesting
physics.
The $2S$ single particle wave functions  
$\langle {\bf r} |i\rangle$ oscillate
rapidly. Thus the transition intensity to state $i$, governed by
the overlap factor $F^{i}_{DF}=|\langle i|0\rangle|^{2}$,
is most sensitive to the value of
$\psi({\bf r})=\langle {\bf r} 
|0\rangle=\sqrt{n({r})/N}$ at the state's classical 
turning points. At a given laser frequency, the excitation is resonant
with all states with motional energy $\varepsilon=2h\nu-E_{1S-2S}$.
This suggests the excitation rate
is proportional to the integral of the
condensate density in a shell at the equipotential surface defined by 
the classical turning points
of $2S$ states with  motional energy $\varepsilon$.

For a  spherically symmetric trap, we can formally show
this  by making WKB and static phase approximations \cite{jab45,aki82}
 - a technique  which has recently been 
applied  to describe $s$-wave  collision
photoassociation spectra \cite{jul96} and quasiparticle excitation in 
a condensate \cite{cgs98}. 
One uses a WKB 
expression for the $2S$ eigenstate.
Then, because of the slow spatial
variation of the condensate wave function,
the Doppler-free overlap factor  only
depends on the condensate wavefunction and the $1S$ and $2S$ potentials
where the phase of the upper state is
stationary. This yields
\begin{eqnarray}
F^{i}_{DF}&=&|\langle i|0\rangle|^2 \approx 4\pi 
\left|R_{i}\sqrt{n({R_{i}})\over N}\right|^{2}/D, 
\label{fc}
\end{eqnarray}
where $R_{i}$ is the Condon point, or the radius where the 
local wave vector of the excited state 
($k_{2S}=\sqrt{{2m \over \hbar^{2}}[\varepsilon_{i}-V_{2S}^{eff}(r)]}$)
vanishes. $R_{i}$ is equivalent to the classical turning point for state 
$i$, and is defined through    
\begin{equation}
\varepsilon_{i}=V_{2S}^{eff}(R_{i}). \label{condon}
\end{equation}
Also, in the limit that we can neglect the slow spatial
variation of the BEC wave function, 
$D\approx d V_{2S}^{eff}(r)/dr|_{R_{i}}\equiv V'^{eff}_{2S}(R_{i})$ 
is the slope of the
effective $2S$  potential at the Condon point.

Using Eq.\ \ref{fc}, the Fermi's Golden Rule expression for
the spectrum (Eq.\ \ref{overlapint}) becomes
\begin{eqnarray}
S_{DF}¥(2h\nu) &=&{N \pi \hbar \Omega_{DF}^2 \over 2}
\sum_{i}  {4\pi \left|R_{i}\sqrt{n({R_{i}})\over N}\right|^{2}\over V'^{eff}_{2S}
(R_{i})}\,  
\delta(2h\nu-E_{1S-2S} - \varepsilon_i + \mu).
\label{temp} 
\end{eqnarray}
The Doppler-free excitation  field and the BEC wave function are
spherically symmetric, so
only $2S$ motional states
with zero angular momentum are excited. 
This implies that in the limit of closely spaced levels, 
$\Sigma_{i}\rightarrow \int d\varepsilon$ in Eq.\ \ref{temp}.
Using Eq.\ \ref{condon} we can change variables:
$\int d\varepsilon = \int dR\, V'^{eff}_{2S}
(R)$ and $\delta\left(2h\nu-E_{1S-2S} - \varepsilon +\mu\right) 
=\delta\left(2h\nu-E_{1S-2S} - {4\pi\hbar^2 \delta a \, n({ 
R})\over m}\right)$, where $\delta a = a_{1S-2S}-a_{1S-1S}$.
This yields
\begin{eqnarray}
S_{DF}(2h\nu) &=&{\pi \hbar \Omega_{DF}^2 \over 2}
\int4\pi dr\, r^{2}n({r})
\, \delta\left(2h\nu-E_{1S-2S} - {4\pi\hbar^2 \delta a \, n({ 
r})\over m}\right). \label{radialintegral}
\end{eqnarray}

Using the probabilistic interpretation of
$|\psi({\bf r}_{i})|^{2}$ (Sec.\ \ref{condsec}),
one can interpret Eq.\ \ref{radialintegral} in the following way.
When a $2S$ excitation
is detected at a given frequency, it 
records the fact that a $1S$ atom was found at a position which
had a $1S$ density which brought that atom into resonance with the
laser. The rate of excitation
is proportional to the probability of finding a condensate atom in a region
with the correct density. This is a local density description of the 
spectrum, and it is justified by the slow spatial variation of the 
condensate wave function.

For a Thomas-Fermi wave function 
in a three dimensional harmonic
trap, Eq.\ \ref{radialintegral} reduces to 
\begin{eqnarray}
S_{DF}¥(2h\nu) &=& 
{15 \pi \hbar \Omega_{DF}^2  N\over 8}
{ (E_{1S-2S}-2h\nu)
\over
\left(2h\, \delta \nu_{max}\right)^2
}
\left[1-{2h\nu-E_{1S-2S}\over 
2h\, \delta \nu_{max}}
\right]^{1/2} \label{reduce}
\end{eqnarray}
for $2h\, \delta \nu_{max} <2h\nu-E_{1S-2S} <0$,
 and otherwise $S_{DF}¥(2h\nu)=0$.
 Here, $2h\, \delta \nu_{max} = 
4\pi\hbar^2 \delta a \, n(0)/m.$

Figure \ref{overlap} shows that for a spherically 
symmetric trap, Eq.\ \ref{reduce} 
agrees with the spectrum calculated directly with
Fermi's Golden Rule (Eq.\ \ref{overlapint}) using simple harmonic 
oscillator wave functions. 
For a trap which has a weak confinement axis, such as the MIT hydrogen
trap \cite{fkw98,kfw98}, 
discrete transitions in the spectrum are too closely spaced to be  
resolved.
The envelope given by Eq.\ \ref{reduce}, however,
shows no dependence on the trap frequencies or the
symmetry (or lack thereof) of the harmonic trap.

\begin{center}
\epsfig{file=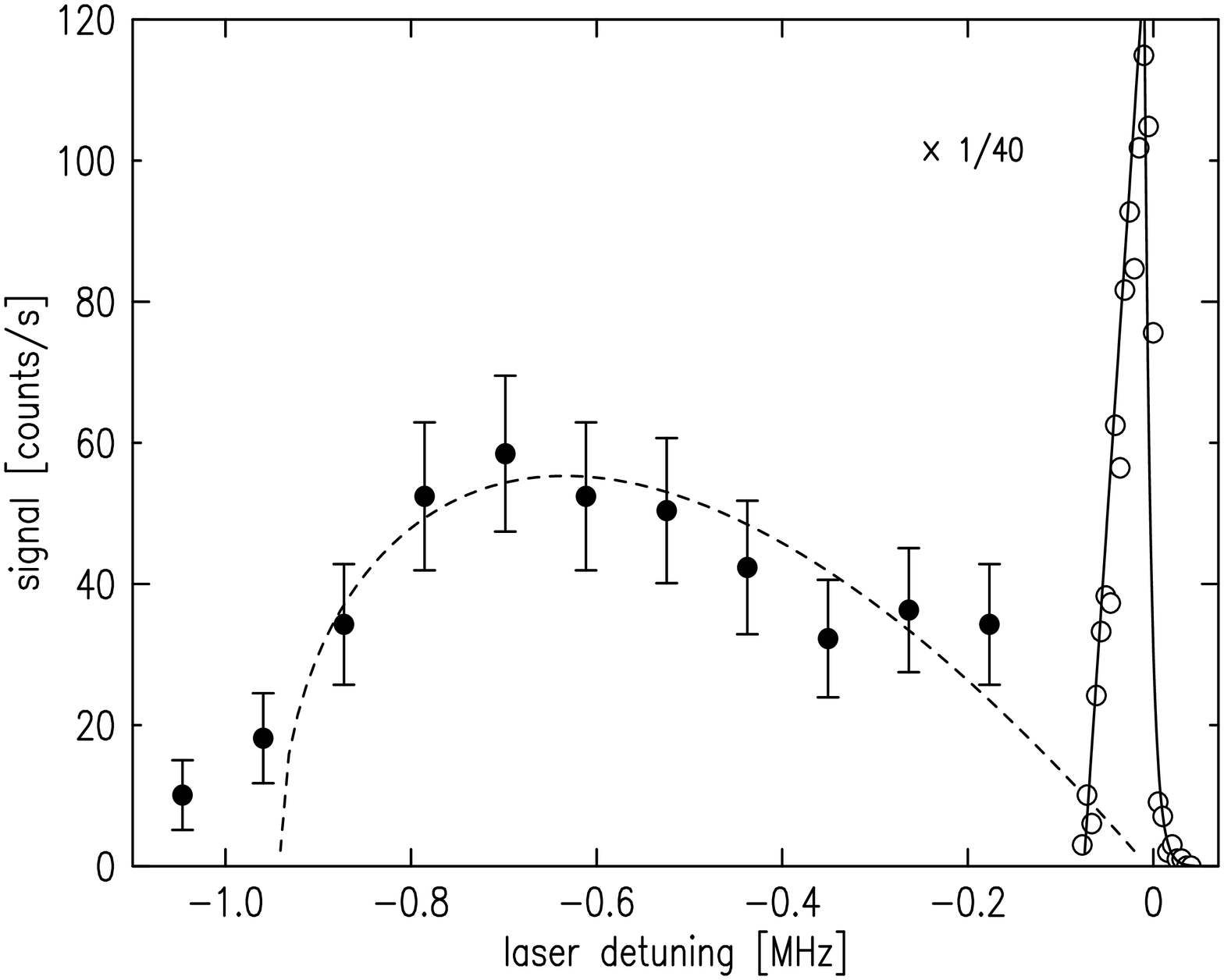, width=3in}

\begin{figure}
\caption
{Doppler-free spectrum of a condensate: comparison of theory and  
experiment (from [1]). 
The narrow feature near zero detuning
is the spectral contribution from the noncondensed atoms (shown 
$\times 1/40$). The broad 
feature is the
spectrum of the condensate. The dashed curve is Eq.\ 
\ref{reduce}, which comes from the integral over the BEC density 
distribution (Eq.\ \ref{radialintegral})
assuming a Thomas-Fermi density distribution for a harmonic trap.
}
\label{condspec}
\end{figure}

\end{center}

Theory and experimental data are compared in
Fig.\ \ref{condspec}. Although the statistical error bars for the
data are large due to the small number
of counted photons, 
the theoretical BEC spectrum for a
condensate at $T=0$
fits the data reasonably well.
The deviations may indicate nonzero temperature effects or 
reflect experimental noise.
The smoothing of the cutoff
at large detuning may be due to
shot to shot variation in the peak
condensate density for the 10 atom trapping cycles which contribute
to this composite spectrum.
Also, at low detuning the BEC spectrum
is  affected by the wing of the Doppler-free
line for the noncondensed atoms.

Using this theory,  from 
the peak shift in the spectrum, the trap oscillation frequencies,
 and knowledge  of $a_{1S-1S}$ and $a_{1S-2S}$,
one can calculate the number of atoms in the condensate.
Assuming the experimental value of $a_{1S-2S}$,
the result is 
larger  than the number determined from a model
of the BEC lifetime and loss rates, which
is discussed in  \cite{wkf99}. The uncertainties are large for these 
results, but the disagreement
could be due to error in  the experimental value of $a_{1S-2S}$, 
uncertainty
in the gas temperature or trap and laser parameters, or thermodynamic conditions
in the trapped gas
which are different than assumed by the theories.
For example, we have implicitly assumed  local
spatial coherence ($g^{(2)}(0)=1$) \cite{kmi97}
in our form of the BEC wave function (Eq.\ \ref{define}). 
It has not yet been experimentally
verified that the hydrogen condensate is coherent.

\subsection{Doppler-Sensitive 1S-2S Spectrum}
In contrast to the Doppler-free excitation spectrum,
the Doppler-sensitive  spectrum in principle reflects the
finite momentum spread in the condensate as well as the mean field effects.
The relevant momentum spread is given by the uncertainty principle and is 
$\sim \hbar /\delta 
z$ where
$\delta z \approx 5$~mm is the length of the condensate along the 
laser propagation 
axis.
However, in the hydrogen experiment the cold collision frequency shift
($\sim 1$~MHz) 
dominates over the Doppler-broadening in the spectrum ($\hbar k_{0} /2\pi 
m\,\delta 
z\approx  100$~Hz.)
We can thus neglect Doppler-broadening, which is equivalent to 
neglecting the spatial variation of the BEC wave function in any
transition matrix elements.
In this regime it is possible to modify the derivation of the WKB
and static phase
approximations \cite{jab45,aki82,jul96,cgs98} 
to calculate the Doppler-sensitive 
spectrum. 

We rewrite the Doppler-sensitive Rabi frequency (Eq.\ \ref{dsrabi}) as
\begin{eqnarray}
\Omega_{DS}({\bf r})&=&\Omega_{DS}\left({\rm e}^{ik_{0}z} +{\rm 
e}^{-ik_{0}z}\right) \nonumber \\
&=&2 \Omega_{DS}\sum_{l\, even}\sqrt{4\pi 
(2l+1)}i^{l}j_{l}(k_{0}r)Y_{l}^{m=0}(\theta,\phi), \label{jackson}
\end{eqnarray}
where $j_{l}(k_{0}r)$  is the spherical Bessel function of
order $l$, and $Y_{l}^{m}(\theta,\phi)$ is a spherical harmonic. 
This shows that the Doppler-sensitive laser Hamiltonian can excite atoms
to $2S$ motional states with any even value of angular momentum, but with $m=0$.

Transitions are to levels with motional energy 
$\sim \hbar^{2}k_{0}^{2}/2m $ above the bottom of the $2S$
potential, so we label levels by $\Delta$, their energy deviation 
from this value.
For simplicity, we consider a spherically symmetric trap.
This allows us to write
a general expression for the $2S$ wave functions
$\psi_{\Delta,l}=Y_{l}^{m=0}(\theta,\phi)u_{\Delta,l}(r)/r$
where 
$u_{\Delta,l}(r)/r$ satisfies
\begin{eqnarray}
\left[-{\hbar^{2} \over 2m} {d^{2} \over dr^{2}} 
+ {\hbar^{2} \over 2m}{l(l+1) \over r^{2}} + V_{2S}^{eff}(r)\right]u_{\Delta,l}(r)
=\left(E_{1S-2S} + {\hbar^{2} k_{0}^{2} \over 2m} +\Delta \right)u_{\Delta,l}(r).
\label{radse}
\end{eqnarray}

Using Eq.\ \ref{overlapint}, the spectrum  is
\begin{eqnarray}
S_{DS}¥(2h\nu) &=&{N \pi \hbar \over 2}
\sum_{\Delta,l} |\langle \psi_{\Delta,l}|\Omega_{DS}({\bf r})|0\rangle|^2\,  
\delta\left(2h\nu-E_{1S-2S} -  {\hbar^{2} k_{0}^{2} \over 2m} -\Delta + 
\mu\right). 
\label{dsoverlapinttext} 
\end{eqnarray}
Using Eq.\ \ref{jackson}, the overlap integral we must evaluate is
\begin{eqnarray}
\langle \psi_{\Delta,l}|
{\rm e}^{ik_{0}z} +{\rm e}^{-ik_{0}z}
|0\rangle  
&=& \int dr\, r u_{\Delta,l}(r) 2\sqrt{4\pi 
(2l+1)}i^{l}j_{l}(k_{0}r) \sqrt{{n(r) \over N}} \label{step1text}
\end{eqnarray}
for $l$ even, and $0$ otherwise.
Because $\sqrt{n(r)}$  varies slowly, one can find an 
approximate expression for  
this matrix element.
Appendix \ref{dopfreeapp} gives the details of this derivation and uses 
the result to 
reformulate
Eq. \ref{dsoverlapinttext} as
\begin{eqnarray}
S_{DS}¥(2h\nu) 
&\approx&{ \pi \hbar \Omega_{DS}^{2}}
\sum_{\Delta}
4\pi R^{2}_{\Delta} {n(R_{\Delta}) \over V'^{eff}_{2S}(R_{\Delta})} 
\, \delta(2h\nu-E_{1S-2S} -  {\hbar^{2} k_{0}^{2} \over 2m} -\Delta + 
\mu). 
\label{dsoverlapintreducetext} 
\end{eqnarray}

\begin{center}

\begin{figure}
\epsfig{file=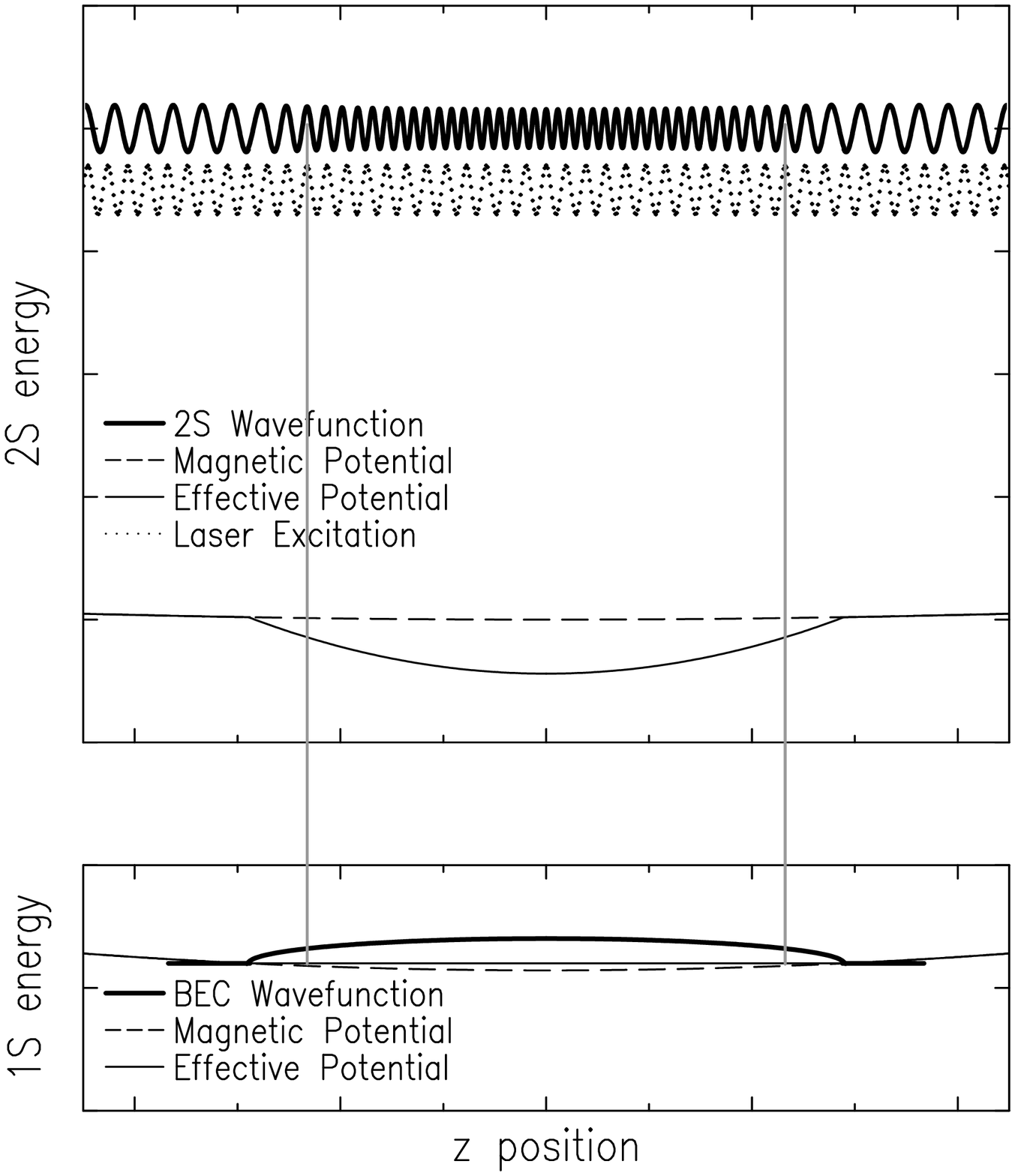, width=3in}

\caption{Effective
potentials, wave functions and the laser field for
Doppler-sensitive excitation of condensate atoms. 
The spatial period 
of the $2S$ wave function, the laser wavelength, and
the vertical axes for the potentials  are not to scale. The vertical
axes for the wave functions and laser field are arbitrary.
In the overlap integral for the
transition matrix element (Eq.\ \ref{step1text}), the
only nonzero contribution comes from
the region where the spatial period  of the $2S$ wave function matches
the wavelength of the laser field. This is indicated by the locations
of the light vertical lines. As the laser frequency is changed, the region of
wavelength match moves. }
\label{veffdsfig}
\end{figure}

\end{center}

The matrix element (Eq.\ \ref {step1text}) gets it's main contribution at
$R_{\Delta}¥$
where the classical wave vector of the WKB approximation for 
$u_{\Delta,l}$ equals  the classical wave vector of the WKB approximation for
$j_{l}$. In effect, $R_{\Delta}$ is the point where
 the spatial period of the wave function matches
the wavelength of the laser field, $2\pi/k_{0}$ (see Fig.\ \ref{veffdsfig}).
This leads to a definition for $R_{\Delta}$
\begin{equation}
\Delta =V_{2S}^{eff}(R_{\Delta}), \label{dscondontext}
\end{equation}
which is identical
to Eq.\ \ref{condon},
the definition of the Condon point from the calculation of the Doppler-free spectrum.
Because the transition is localized in this way, the matrix element 
(Eq.\ \ref {step1text}) is 
proportional to $\sqrt{n(R_{\Delta})}$, 
as is evident in Eq.\ \ref{dsoverlapintreducetext}.

Using Eq.\ 
\ref{dscondontext},
we can replace the sum in Eq.\ \ref{dsoverlapintreducetext} with an integral and change
variables, $\Sigma_{\Delta}\rightarrow \int d\Delta = \int dR\, V'^{eff}_{2S}
(R)$. This yields the Doppler-sensitive
lineshape 
\begin{eqnarray}
S_{DS}¥(2h\nu) &\approx&{\pi \hbar \Omega_{DS}^{2}}
\int
4\pi dr\, r^{2} n(r) 
\, \delta \left(2h\nu-E_{1S-2S} -  {\hbar^{2} k_{0}^{2} \over 2m} 
-{4\pi\hbar^2 \delta a \, n({r})\over m}\right).
\label{dsreflection}
\end{eqnarray}
The Doppler-sensitive condensate spectrum has the same shape as
the Doppler-free spectrum, but it is shifted to the blue by 
photon momentum-recoil. Because 
$\Omega_{DS}=\Omega_{DF}/2$,
the Doppler-sensitive spectrum is half as intense as the 
Doppler-free. 

In  \cite{wkf99}, experimental data are compared with Eq.\ \ref{dsreflection},
and the agreement is good.

\section{Other Applications of the Formalism
}
\subsection{Other Atomic Systems and Excitation Schemes}
\label{otherschemes}
We have specifically considered $1S$-$2S$ spectroscopy of hydrogen, but
the formalism is more general.
For instance, 
if the ground-excited 
state
interaction were repulsive, this would simply modify the effective $2S$
potential (Eq.\ \ref{2seff}) and the form of the motional
states excited by the laser would change. Equations \ref{radialintegral} and 
\ref{dsreflection} 
would still be accurate for two-photon excitation to
a different electronic state when the mean field interaction dominates the 
spectrum. 

In the recently observed  rf hyperfine spectrum of a rubidium 
condensate \cite{bhe99}, the lineshape is determined by mean field
energy and the different magnetic potentials felt by atoms in the initial and 
final states. 
The theory presented here can be modified to describe this situation as well.

For Bragg diffraction or spectroscopy 
as performed in \cite{kdh99,sic99}, atoms remain 
in the same internal state after excitation. 
Particle exchange 
symmetry of the wave function modifies the mean field
interaction energy of the excited atoms
with the atoms remaining in the condensate.
In terms of the hydrogen levels,  
$1S$, $F=1$, $m_{f}=1$ atoms not in the
condensate  experience a potential of 
$8\pi \hbar^2 a_{1S-1S}n_{1S}({\bf r})/m$.
This is to be compared with the mean field potential
of $4\pi \hbar^2 a_{1S-2S}n_{1S}({\bf r})/m$ experienced
by $2S$ particles excited out of the condensate and
$4\pi \hbar^2 a_{1S-1S}n_{1S}({\bf r})/m$  experienced by
$1S$ atoms in the
condensate.
In appendix \ref{excbectrap}, the point in the derivation where
the difference arises is indicated.

\subsection{Doppler Broadening in the Doppler-Sensitive  Spectrum}

 To derive the Doppler-sensitive $1S$-$2S$ spectrum,
we  neglected the variation of the condensate
wave function, which is equivalent to neglecting the atomic 
momentum spread. This is well justified for the hydrogen experiment. The 
effect of small but nonnegligible momentum is discussed at the end of
 appendix \ref{dopfreeapp}. Now
we briefly describe the Doppler-sensitive lineshape when Doppler-broadening 
is dominant. The 
lineshape turns out to be similar to that which was
seen with Bragg spectroscopy of a Na condensate\cite{sic99}.

When the mean 
field potential can be neglected,
the $2S$ motional wave functions are approximately   those of the simple 
harmonic oscillator potential produced by the magnetic trap alone.
Because the spatial extent for these motional states is large compared to 
$\delta z$,
in the region of the condensate the wave functions
can be represented as plane waves 
momentum eigenstates $|{\bf p}\rangle$ \cite{ckl99}. The spectrum becomes
\begin{eqnarray}
S_{DS}¥(2h\nu) &\approx&{N \pi \hbar \over 2}
\sum_{{\bf p}} |\langle {\bf p}|\Omega_{DS}({\bf r})|0\rangle|^2\,  
\delta\left(2h\nu-E_{1S-2S} -  {p^{2} \over 2m}  + 
\mu\right)\nonumber \\
&=& {N \pi \hbar \Omega_{DS}^{2}\over 2}{2\over (2\pi \hbar)^{3}¥
}\int_{p_{z}¥>0} d^{3}p\,
|A({\bf p}-\hbar k_{0}{\bf{\hat z}})|^{2}\delta\left(2h\nu-E_{1S-2S} -  
{p^{2} \over 2m}  + 
\mu\right).
\label{fermimom} 
\end{eqnarray}
The Fourier transform of the condensate 
wave function,
$
A({\bf p})=\int d^{3}r\, {\rm e}^{-i{\bf p}\cdot {\bf r}/\hbar}¥ 
\psi_{N}({\bf r}),
$
is nonzero for 
$| p_{x}|$\raisebox{-.6ex}{$\stackrel{<}\sim$}$\hbar/|\delta x|$,  
 $| p_{y}|$\raisebox{-.6ex}{$\stackrel{<}\sim$}$\hbar/|\delta y|$, and
 $| p_{z}|$\raisebox{-.6ex}{$\stackrel{<}\sim$}$\hbar/|\delta z|$.

The excited states have  $p_{z}\approx \hbar k_{0}$, 
so we define $\delta p=p_{z}-\hbar k_{0}$. Because
the laser 
wavelength is  small compared to the spatial extent
of the condensate,
${p^{2} /2m}\approx \hbar^{2} k_{0}^{2} / 2m +\hbar k_{0} \delta 
p /m$ and the spectrum reduces to
\begin{eqnarray}
S(2h\nu)&\approx& {N \pi m \Omega_{DS}^{2}\over k_{0}(2\pi \hbar)^{3}¥}
\int dp_{x}¥dp_{y}¥
|A( p_{x}¥{\bf{\hat x}}+p_{y}¥{\bf{\hat y}}+\delta p(\nu){\bf {\hat z}})|^{2},
\label{end} 
\end{eqnarray}
where
\begin{eqnarray}
{\hbar k_{0}¥\delta p(\nu) \over m} = 2h\nu-E_{1S-2S} -  
{\hbar^{2}k_{0}^{2} \over 2m}  + 
\mu
\end{eqnarray}
defines the momentum class that is Doppler shifted into resonance.
The spectrum is centered at $2h\nu=E_{1S-2S} +  
{\hbar^{2}k_{0}^{2} / 2m}-\mu$, and the
lineshape  depends on the  orientation of the condensate wave function
with respect to the laser propagation axis. 

For a Thomas-Fermi wave function in a spherically symmetric harmonic trap
$|A({\bf p})|^{2}\sim |j_{2}(pr_{0}/\hbar)/(pr_{0}/\hbar)^2|^{2}$ \cite{bpe96},
where $r_{0}=\sqrt{2n(0){\tilde 
U}/mw^{2}}$.
Numerical evaluation of the integral over $p_x$ and $p_y$ shows that the lineshape is 
approximately given by  the power spectrum of the wave function's spatial variation 
along $z$, $S(2h\nu) \propto |A(\delta p(\nu){\bf {\hat z}})|^{2}$.

In recent
experiments with small angle light
scattering \cite{scg99},  the momentum imparted to atoms is small compared
to $\sqrt{2}mc_{s}$, where $c_{s}=\sqrt{\mu/m}$ is the speed of 
Bogoliubov sound. 
In this case  one can excite quasiparticles in the 
condensate as opposed to free particles. The theory described 
in this article only
treats free particle excitation, but
Bogoliubov formalism, combined with  WKB and
static phase approximations, has been  used to describe
the spectrum for quasiparticle excitation \cite{cgs98}.

\section{Discussion}
To make the problem analytically tractable,
we have only  derived the BEC spectrum for the specific case of a spherically
symmetric trap. The trap shape does not appear in the
final expressions (Eq.\ \ref{radialintegral} and \ref{dsreflection}), 
however, 
and  with reasonable confidence
we can extend the results to any geometry. In the experiment,
the trap aspect ratio is as large as 400 to 1, but the data agrees well
with this theory. The physical picture of the transition occurring
at the classical turning points, and the probabilistic 
or local density interpretation
of the spectrum also support the generalization of 
Eq.\ \ref{radialintegral} and \ref{dsreflection} to
\begin{eqnarray}
S_{DF}¥(2h \nu) &=& 
{\pi \hbar \Omega^{2}_{DF}¥ \over 2}\int d^3r\, n_{1S}({\bf r})
\delta\left(2h\nu-
E_{1S-2S}¥ -{4\pi\hbar^2 \delta a \, n_{1S}
({\bf r})\over 
m}\right) \label{dfapprox}\\
S_{DS}¥(2h \nu) &=& 
{\pi \hbar \Omega^{2}_{DS}}\int d^3r\, n_{1S}({\bf r})
\delta\left(2h\nu-
E_{1S-2S}-{\hbar^{2}k_{0}^{2}\over 2m}-{4\pi\hbar^2 \delta a \, n_{1S}
({\bf r})\over 
m}\right) 
.  \label{localapprox}
\end{eqnarray}

 Equations \ref{dfapprox} and \ref{localapprox}  
take ${4\pi \hbar^2 \delta a \,  n_{1S}({\bf r})/ m}$ 
as a local  shift of the transition
frequency and ascribe
the excitation to a
small region in space where the laser is resonant. 
This approach is similar to a quasistatic
approximation in standard spectral lineshape theory \cite{aki82} which
neglects the atomic motion and averages over the distribution of 
interparticle spacings to find the spectrum. Atom pairs at different separations
experience different frequency shifts due to  atom-atom
interactions. This broadens the  line.

There are important differences between the theory presented here
and the quasistatic approximation, however. For the standard quasistatic
treatment to be valid,  the lifetime of the
excited state should be shorter than a collision time \cite{bar62}.
For a condensate, the classical concept of a collision time
is inapplicable.
We have shown that Eq.\ \ref{dfapprox} and \ref{localapprox} result 
from a different
approximation: neglecting 
the slow spatial variation of the BEC wave function. Also, for the
condensate spectrum, one integrates over atom position in the effective
potential, as opposed to integrating over the distribution of 
atom-atom separations.
Finally, the BEC spectral broadening
is  homogeneous, which is not  normally the case when making the 
quasistatic approximation.

It is interesting that although
the atoms in the condensate are delocalized over a region
in which the density varies from it's maximum value to zero, the rapid
oscillation of the excited state wave function essentially localizes the
transition (Eq.\ \ref{condon} and \ref{dscondon}). In this way, the excitation 
probes the condensate
wave function spatially.

The description of the BEC spectrum developed here has provided insight into the
excitation process and it is  general. 
We have 
shown that the formalism of  transitions between bound states of the effective 
potentials can be used when either the mean field or Doppler 
broadening dominates. It can describe a variety of excitation 
schemes such as two-photon Doppler-free or Doppler sensitive 
spectroscopy to an excited electronic state, or Bragg
diffraction which leaves the atom in the ground state.

\bigskip
\noindent
{Acknowledgments}

We thank 
D. Kleppner for comments on this manuscript
and, along with T. Greytak, for guidance during the course of 
this study. Discussions of the hydrogen experimental results with 
D. Fried, D. Landhuis, S. Moss, and in particular L. Willmann
inspired much of this theoretical 
work and provided valuable feedback. Thoughtful contributions from 
W. Ketterle, L. Levitov, M. Oktel, and  P. Julienne, 
and discussions with  E. Tiesinga regarding
 the proper form of the collision
Hamiltonian, Eq.\ \ref{AinteractionH}, are gratefully 
acknowledged.
Financial support was provided by the National Science Foundation and 
the Office of Naval Research.

\appendix

\section{Energy Functional for the System after Laser Excitation}
\label{excbectrap}
In this appendix we derive Eq.\ \ref{some2SBECenergy},
the energy functional for the system after excitation which is
minimized to find the $2S$  wave functions.

The Hamiltonian and the excited state vector,
$|\Phi_{q,i}\rangle$, are defined in Eq.\ \ref{Htrap} 
and  \ref{excitedbra}. The symmetry operator
is explicitly written as
$
\hat{\cal S}={\scriptsize \left(\begin{array}{c}
N \\ q
\end{array} \right) }^{-1/2}\sum_{P}P
$, where the sum runs over the
${\scriptsize \left(\begin{array}{c}
N \\ q
\end{array} \right) }=
{N! \over q!(N-q)!}$ distinct particle label permutations $P$.
The energy functional  for $N-q$ $1S$ condensate atoms and
$q$ $2S$ atoms in state $i$ is
\begin{eqnarray}
\langle \Phi_{q,i}|H |\Phi_{q,i}\rangle 
&=&\langle \Phi_{q,i}|
\sum_{j=1}^N\left({{ p}_j^2 \over 2m} + V({\bf { r}}_j) + 
H^{int}_j\right) + H^{coll}
|\Phi_{q,i}\rangle \nonumber \\
&=&(N-q)\langle 1S,0|
\left({{ p}^2 \over 2m} + V({\bf { r}})+ H^{int}\right) 
|1S,0\rangle \nonumber \\
& & + q\langle 2S,i|\left({{ p}^2 \over 2m} + V({\bf 
{ r}})+ H^{int}\right) 
|2S,i\rangle + \langle \Phi_{q,i}|H^{coll} 
|\Phi_{q,i}\rangle. \label{temp100}
\end{eqnarray}

We  evaluate the interaction term,
\begin{eqnarray}
\langle \Phi_{q,i}|H^{coll} |\Phi_{q,i}\rangle 
&=&\langle 2S;...;1S;..|\hat{\cal S}H^{coll}
\hat{\cal S}|2S;...;1S;...\rangle \nonumber \\
&=&\langle 2S;...;1S;...|H^{coll}
\hat{\cal S}\hat{\cal S}|2S;...;1S;...\rangle \nonumber \\
&=&\langle 2S,...;1S;...|H^{coll}
\hat{\cal S}|2S;...;1S;...\rangle{\scriptsize \left(\begin{array}{c}
N \\ q
\end{array} \right) }^{1/2} \nonumber \\
&=&\langle 2S;...;1S;...|H^{coll}\sum_{P}P|2S;...;1S;...\rangle,
\label{temp10}
\end{eqnarray}
where we have  used $[H^{coll},\hat{\cal S}]=0$ and  
$\hat{\cal S}\hat{\cal S}=
\hat{\cal S}{\scriptsize \left(\begin{array}{c}
N \\ q
\end{array} \right) }^{1/2}$. 
Of the $N(N-1)/2$ terms in $H^{coll}$ (Eq. 
\ref{AinteractionH}), 
$(N-q)(N-q-1)/2$ of them
result in a $1S$-$1S$ interaction,
$(N-q)q$ of them
result in a $1S$-$2S$ interaction, and the
rest result in a $2S$-$2S$ interaction which we can neglect. 
For the $1S$-$1S$ terms, only the identity permutation contributes.
For the $1S$-$2S$ terms two permutations contribute - the identity
and switching the labels on the two interacting particles. The expectation
value of $H^{coll}$ thus reduces to
\begin{eqnarray}
{2\pi \hbar^2 \over m}(N-q)(N-q-1) a_{1S-1S}\langle 0;0|
\delta ({\bf { r}}_1 - {\bf { r}}_2)
|0;0\rangle +{4\pi \hbar^2 \over m}q(N-q) a_{1S-2S}\langle i;0|
\delta ({\bf { r}}_1 - {\bf { r}}_2)
|i;0\rangle. \label{collterm}
\end{eqnarray}
As mentioned in Sec. \ref{otherschemes}, Eq. 
\ref{temp10} would be modified  for 
Bragg diffraction or spectroscopy  as performed in \cite{kdh99,sic99}
because the internal state is unchanged during laser excitation. 
We do not explicitly treat this situation because it is not central to this 
study.

Inserting Eq. \ref{collterm} into Eq.\ \ref{temp100}, we find the
energy functional is 
\begin{eqnarray}
\langle \Phi_{q,i}|H |\Phi_{q,i}\rangle &=&E'_{0} \nonumber \\
& & + q\, \langle 2S, i|
\left[ H^{int} +{{ p}^2 \over 2m}  + V({\bf { r}})+  
{4\pi\hbar^2a_{1S-2S} \over m} \, n_{N-q}({\bf { r}}) \right]
|2S,i\rangle \nonumber \\
&=&E'_{0} + q(E_{1S-2S} + \varepsilon_i),
\end{eqnarray}
where 
\begin{eqnarray}
E'_{0}
&=&(N-q)\langle 1S,0|
\left({{ p}^2 \over 2m} + V({\bf { r}})+ H^{int}\right) 
|1S,0\rangle \nonumber \\
& & + {2\pi \hbar^2 \over m}(N-q)(N-q-1) a_{1S-1S}\langle 
0;0|
\delta ({\bf { r}}_1 - {\bf { r}}_2)
|0;0\rangle, \label{eprime}
\end{eqnarray}
is the energy for $N-q$ isolated
$1S$ condensate atoms, and
$\varepsilon_i=\langle i|
\left[{{ p}^2 \over 2m}  + V_{2S}^{eff}({\bf r}) \right]
|i\rangle$.
The density in the condensate for $N-q$ condensate atoms ($q \ll N$)  is
$
n_{N-q}({\bf r})=(N-q)\langle 0|
\delta ({\bf { r}}_1 - {\bf  r})
|0\rangle .  
$

\section{WKB and Static Phase  Approximations for the Doppler-Sensitive
BEC Spectrum}
\label{dopfreeapp}
In this appendix we calculate the
Doppler-sensitive overlap integral, Eq.\ \ref{step1text},
and simplify Eq.\ \ref{dsoverlapinttext}.
The derivation is   similar to the treatment of \cite{jul96,cgs98}.

The overlap integral we must evaluate is
\begin{eqnarray}
I_{\Delta,l} &=&\langle \psi_{\Delta,l}|
{\rm e}^{ik_{0}z} +{\rm e}^{-ik_{0}z}
|0\rangle  \nonumber \\
&=& \int dr\, r u_{\Delta,l}(r) 2\sqrt{4\pi 
(2l+1)}i^{l}j_{l}(k_{0}r) \sqrt{{n(r) \over N}} \label{step1}
\end{eqnarray}
for $l$ even and $0$ otherwise.

Because $u_{\Delta,l}$ and $j_{l}$ are rapidly varying
compared to $\sqrt{n}$ it is useful to
express $u_{\Delta,l}$ and $j_{l}$ in phase-amplitude
form through a WKB approximation. We define the local wave vectors
for $u_{\Delta,l}$ and $j_{l}$
\begin{eqnarray}
k_{u}(\Delta,l,r)&=&\left[k_{0}^{2} -{l(l+1) \over r^{2}} - {2m \over 
\hbar^{2}}\left(V_{2S}^{eff}(r) -\Delta \right)\right]^{1/2}, \label{ku} \\
k_{j}(k_{0},l,r)&=&\left[k_{0}^{2} -{l(l+1) \over r^{2}}\right]^{1/2}. \label{kj}
\end{eqnarray}
Then, in the classically allowed region 
\begin{eqnarray}
u_{\Delta,l}(r)&\approx&
{1\over \sqrt{k_{u}(\Delta,l,r)}}\left({2m \over \pi \hbar^{2}}\right)^{1/2}
{\rm sin}\beta_{u}(\Delta,l,r), \label{u} \\
j_{l}(k_{0}r)&\approx&
{1\over r\sqrt{k_{0}k_{j}(k_{0},l,r)}}
{\rm sin}\beta_{j}(k_{0},l,r), \label{j}
\end{eqnarray}
where 
\begin{eqnarray}
\beta_{u}(\Delta,l,r)&=&\int_{R_{T}^{\Delta,l}}^{r}dr'\, k_{u}(\Delta,l,r') - 
\pi/4, 
\label{ub} \\
\beta_{j}(k_{0},l,r)&=&\int_{R_{T}^{k_{0},l}}^{r}dr'\, k_{j}(k_{0},l,r') 
-\pi/4 
\label{uj}
\end{eqnarray}
are the phases. The inner turning points against the centrifugal 
barriers are denoted by $R_{T}$.
Note that the approximations are good for $(k_{0}r)^{2}¥ > l(l+1)$. 
For $(k_{0}r)^{2}¥ < l(l+1)$, neglecting the small $V_{2S}^{eff}$ and 
$\Delta$, the functions behave as  damped 
exponentials. The outer turning points are of no concern to the 
calculation.

Now we write
\begin{eqnarray}
I_{\Delta,l\, even} &=&-2\sqrt{4\pi (2l+1)}
\int dr\,  \sqrt{{n(r)\over N}} 
{{\rm sin}\beta_{u}(\Delta,l,r)\over \sqrt{k_{u}(\Delta,l,r)}}\left({2m \over \pi \hbar^{2}}\right)^{1/2}
{{\rm sin}\beta_{j}(k_{0},l,r)\over \sqrt{k_{0}k_{j}(k_{0},l,r)}}
 \nonumber \\
&\approx&-\sqrt{4\pi (2l+1)}
 \left({2m \over \pi 
\hbar^{2}}\right)^{1/2}
\int dr\,  \sqrt{n(r) \over Nk_{u}(\Delta,l,r)k_{0}k_{j}(k_{0},l,r) }
{\rm cos}\left[\beta_{u}(\Delta,l,r)-\beta_{j}(k_{0},l,r)\right]. 
\nonumber \\
\label{step2}
\end{eqnarray}
We have used the fact that $\sqrt{n(r)}$ varies slowly
and  have dropped rapidly oscillating terms in the integral.

We make the static phase approximation  that the overlap integral 
will only have contributions from the  point 
$R_{\Delta}$ where the difference in the phase factors is stationary.
This point is defined by
$ 
0={d \over dr}\left(\beta_{u}-\beta_{j}\right)|_{R_{\Delta}}=
k_{u}(\Delta,l,R_{\Delta})-k_{j}(k_{0},l,R_{\Delta}),
$
which is equivalent to  an
$l$-independent 
relation defining $R_{\Delta}$ for excitation to states with
energy defect $\Delta$,
\begin{equation}
\Delta =V_{2S}^{eff}(R_{\Delta}). \label{dscondon}
\end{equation}
This is essentially identical to
Eq.\ \ref{condon} from the calculation of the
Doppler-free spectrum.
 
We expand the difference in the phases in a Taylor series around 
$R_{\Delta}$ and write the overlap integral as 
\begin{eqnarray}
I_{\Delta,l\, even}
&\approx&-\sqrt{4\pi (2l+1)} \left({2m \over \pi 
\hbar^{2}}\right)^{1/2}
\sqrt{n(R_{\Delta}) \over Nk_{0}
k^{2}_{j}(k_{0},l,R_{\Delta})} \nonumber \\
&&\times \int_{-\infty}^{\infty} dx\,  
{\rm cos}\left[\beta_{u}(\Delta,l,R_{\Delta})-\beta_{j}(k_{0},l,R_{\Delta}) 
- {mV'^{eff}_{2S}(R_{\Delta}) \over 
2\hbar^{2}k_{j}(k_{0},l,R_{\Delta})}x^{2}\right] \nonumber\\
&=&-\sqrt{16\pi (2l+1)\, n(R_{\Delta}) \over 
Nk_{0}k_{j}(k_{0},l,R_{\Delta}) V'^{eff}_{2S}(R_{\Delta})}
{\rm 
cos}\left[\beta_{u}(\Delta,l,R_{\Delta})-\beta_{j}(k_{0},l,R_{\Delta}) 
- \pi/4 \right].
\label{step3}
\end{eqnarray}
To obtain the last line we have used the Fresnel integral 
$\int_{-\infty}^{\infty} dx\, {\rm cos}(a+bx^{2})=\sqrt{\pi/b}\, 
{\rm cos}(a+{b\over 
|b|}\pi/4)$. Equation \ref{step3} only holds for 
$l(l+1) < (k_{0}R_{\Delta})^{2}$.
For $l(l+1) > (k_{0}R_{\Delta})^{2}$, $I_{\Delta,l\, even}\approx 0$
because $j_{l}(k_{0}r)$  is exponentially
damped at $R_{\Delta}$.

From Eq.\ \ref{dsoverlapinttext} and \ref{step3},
\begin{eqnarray}
S_{DS}¥(2h\nu) &\approx&{\pi \hbar \Omega_{DS}^{2}\over 2}
\sum_{\Delta,l\, even}^{l(l+1) < (k_{0}R_{\Delta})^{2}} 
{16\pi (2l+1) n(R_{\Delta}) 
\over k_{0}k_{j}(k_{0},l,R_{\Delta})V'^{eff}_{2S}(R_{\Delta})}
 \nonumber \\
&&\times{\rm cos}^{2}\left[\beta_{u}(\Delta,l,R_{\Delta})-
\beta_{j}(k_{0},l,R_{\Delta}) - \pi/4\right]
\, \delta\left(2h\nu-E_{1S-2S} -  {\hbar^{2} k_{0}^{2} \over 2m} -\Delta + 
\mu\right)
\end{eqnarray}
We can replace the cos$^{2}$ function with it's average value of 
1/2 because its phase varies rapidly with $l$. Thus
\begin{eqnarray}
&&\sum_{l\, even}^{l(l+1) < (k_{0}R_{\Delta})^{2}}  
{ (2l+1) \over k_{0}k_{j}(k_{0},l,R_{\Delta})}
{\rm cos}^{2}\left[\beta_{u}(\Delta,l,R_{\Delta})
-\beta_{j}(k_{0},l,R_{\Delta}) \right] \nonumber \\
&&\approx{1\over 4}\int_{0}^{l(l+1)=(k_{0}R_{\Delta})^{2}} {dl\, (2l+1) 
\over k_{0}^{2}\sqrt{1-{l(l+1) \over (k_{0}R_{\Delta})^{2}}}} 
\nonumber \\
&&= R^{2}_{\Delta}/2,
\end{eqnarray}
and 
\begin{eqnarray}
S_{DS}¥(2h\nu) 
&=&{ \pi \hbar \Omega_{DS}^{2}}
\sum_{\Delta}
4\pi R^{2}_{\Delta} {n(R_{\Delta}) \over V'^{eff}_{2S}(R_{\Delta})} 
\, \delta(2h\nu-E_{1S-2S} -  {\hbar^{2} k_{0}^{2} \over 2m} -\Delta + 
\mu). 
\label{dsoverlapintreduce} 
\end{eqnarray}

In the derivation given above, we  neglected the variation of the condensate
wave function, which is equivalent to neglecting the atomic 
momentum spread $\sim 
\hbar /\delta r$, where $\delta r$ is the $r$ extent of the 
condensate.
When mean field effects dominate the spectrum, but the atomic momentum is
not completely negligible, the lineshape will deviate from 
Eq.\ \ref{dsreflection}
only for small detunings, 
$\delta \nu$\raisebox{-.6ex}{$\stackrel{<}\sim$}$\hbar k_{0} /2\pi m\,\delta 
r$. One 
can see this from  the overlap integral (Eq.\ 
\ref{step2}) by expressing the condensate wave function in terms of 
the radial Fourier components, 
$
A_{r}¥({ p})=\int dr\, {\rm e}^{-i p r/\hbar}¥ \psi({ r}),
$
to obtain
\begin{eqnarray}
I_{\Delta,l\, even} &=&{-2\sqrt{4\pi (2l+1)}\over 2\pi \hbar}
\int dp\, A_{r}¥({p}) \int dr\,  {\rm e}^{i p r/\hbar} 
{{\rm sin}\beta_{u}(\Delta,l,r)\over \sqrt{k_{u}(\Delta,l,r)}}\left({2m \over \pi \hbar^{2}}\right)^{1/2}
{{\rm sin}\beta_{j}(k_{0},l,r)\over \sqrt{k_{0}k_{j}(k_{0},l,r)}}. \nonumber \\
\label{doppbroad}
\end{eqnarray}

Each momentum component will only contribute to the matrix element
at the point $R_{\Delta,l,p}$ where the total phase under the $r$ integral 
in Eq.\ \ref{doppbroad}
is stationary. This leads to a definition of $R_{\Delta,l,p}$ for each momentum,
$
p 
/\hbar=|k_{u}(\Delta,l,R_{\Delta,l,p})-k_{j}(k_{0},l,R_{\Delta,l,p})|.
$
When $|\Delta| \gg \hbar^{2} 
k_{0}/m\,\delta r$,  $p/\hbar$ is negligible and
this yields the same relation as found by neglecting the curvature
of the BEC wave function (Eq.\ \ref{dscondon}). 
This implies  $S(2h|\delta\nu| \gg \hbar^{2}
k_{0}/m\, \delta r )$ is unaffected by the atomic
momentum. When $|\Delta|$\raisebox{-.6ex}{$\stackrel{<}\sim$}$\hbar^{2} 
k_{0}/m\,\delta r$, the momentum spread in the condensate 
alters $I_{\Delta,l\, even}$. Thus
$S(2h|\delta\nu|$\raisebox{-.6ex}{$\stackrel{<}\sim$}$\hbar^{2} 
k_{0}/m\, \delta r)$ will show some Doppler-broadening
 because of  finite atomic momentum. This effect is 
 negligible for the 
 hydrogen condensate because the cold collision frequency shift 
 ($\sim 1$~MHz) is much greater than the Doppler width  resulting from 
 a 5 mm long condensate wave function ($\hbar k_{0} 
 /2\pi m\,\delta z\sim 100$~Hz).

\end{document}